\documentclass{iau}
\usepackage{graphicx}

\title[Aperiodic variability of NGC 5408 X--1] 
{The aperiodic variability of the Ultraluminous X-ray source in NGC 5408}

\author[M.~D. Caballero-Garcia et~al.]   
{M.~D. Caballero-Garcia $^1$, S.~E. Motta $^1$, T.~M. Belloni $^1$ \and A. Wolter $^2$ }

\affiliation{$^1$ INAF-Osservatorio Astronomico di Brera, Via E. Bianchi 46, \\ I-23807 Merate (LC)
Italy \\ email: {\tt mcaballe@brera.inaf.it } \\[\affilskip]
$^2$ INAF, Osservatorio Astronomico di Brera, via Brera 28, 20121 Milano \\ Italy \\   }

\pubyear{2012}
\volume{290}  
\jname{FEEDING COMPACT OBJECTS: Accretion on All Scales}
\editors{A.C. Editor, B.D. Editor \& C.E. Editor, eds.}
\begin{document}

\maketitle

\begin{abstract}
Ultra-Luminous X-ray sources are accreting black holes that might represent
strong evidence of the Intermediate Mass Black Holes (IMBH), proposed to exist by
theoretical studies but with no firm detection (as a class) so far. We analyze
the best X-ray timing and spectral data from the ULX in NGC~5408 provided by 
{\it XMM-Newton}. The main goal is to study the broad-band noise variability 
of the source. We found an anti-correlation of the fractional root-mean square variability
versus the intensity of the source, similar to black-hole binaries during hard states.

\end{abstract}

\firstsection 
\section{Introduction}

Ultra-Luminous X-ray sources (ULXs) are point like, off-nuclear, extra-galactic sources, with total observed X-ray luminosities
(${\rm L}_{\rm X}{\ge}10^{39}\,{\rm erg}\,{\rm s}^{-1}$) higher than the Eddington luminosity for a stellar-mass black-hole
(${\rm L}_{\rm X}{\approx}10^{38}\,{\rm erg}\,{\rm s}^{-1}$). The true nature of these objects is still open to debate
(\cite{feng11,fender12}) as there is still no unambiguous estimate for the mass of the compact object in these systems.
If the emission is isotropic and the Eddington limit is not violated, ULXs
must be fuelled by accretion onto Intermediate Mass Black Holes (i.e. IMBHs) with masses in the range 
$100-10^{5}\,{\rm M}_{\odot}$ (\cite{colbert99}). It is possible that ULXs appear very luminous due to a 
combination of moderately high mass, mild beaming and mild super-Eddington emission and that ULXs are 
an inhomogeneous population composed of more than one class.

ULXs have been studied intensively during the last decades (see \cite{feng11} for a review). As in the case of Black Hole
Binaries (BHBs), some ULXs undergo spectral transitions from a hard emission-dominated state to a soft emission-dominated state
(see \cite{belloni10a,belloni10b} for a description of the spectral states in BHBs). During the hard state the 
high-energy spectra of ULXs show a power-law spectral shape in the 3-8\,keV spectral range, together with a high-energy 
turn-over at 6-7\,keV, and a {\it soft excess} at low energies (e.g. \cite{kaaret06}). This soft excess can be modelled
by emission coming from the accretion disc and is characterized by a low inner disc temperature of ${\approx}0.2$\,keV,
as expected if the black holes in these sources are indeed IMBHs (\cite{miller03,miller04}). Nevertheless, \cite{soria07}
suggested that some ULXs are consistent with black holes accreting at moderate rate with masses of ${\approx}50-100{\rm M}_{\odot}$.

The study of the timing properties of ULXs represent a promising way to confirm the associations/similarities with this 
class of sources and BHBs. The Power Density Spectra (PDS) of BHBs are usually composed by broad noise components 
and narrow components (quasi periodic oscillations, QPOs). Just a few ULXs show a QPO and their detection has been proven 
to be difficult (\cite{heil09}). In this work we apply a recent tool (from \cite{munoz11}), the Root Mean Square (rms)-Intensity Diagram, which 
has been proved to be useful to map states in BHBs (without the need of any spectral information).

We studied the currently best available dataset from the ULX in the dwarf galaxy NGC~5408, located at a distance of 
${\rm D}=4.8$\,Mpc (\cite{karachentsev02}). NGC~5408 X--1 is very bright, with a peak X-ray luminosity in the (0.3-10\,keV) 
energy range of ${\rm L}_{\rm X}=2{\times}10^{40}\,{\rm erg}\,{\rm s}^{-1}$. \cite{strohmayer09}
found a QPO in its PDS centred at $0.01$\,Hz and inferred a mass
for the black hole in the range of $10^{3}-10^{4}\,{\rm M}_{\odot}$. Recently, \cite{dheeraj12} studied the timing and
spectral properties of NGC~5408 X--1 and have found that the QPO frequency is variable and largely independent on the spectral 
parameters. They suggested that NGC~5408 X--1 is accreting in the {\it saturation} (constancy of the power-law photon index
and disc flux with a further increase in the QPO centroid frequency) {\it regime} a few times observed in BHBs (\cite{vignarca03}). We performed 
timing and spectral studies, but focusing on the evolution of the broad-band noise rather than on the characteristics of the QPO.

\section{Timing and spectral analysis} 

We considered the 6 long (120-130\,ks) high-quality observations performed with {\it XMM-Newton} during 6 years (2006-2011) 
of NGC~5408 X--1. We limited our analysis to the 
EPIC/pn data, due to its higher time resolution and effective area. We refer to Caballero-Garcia et~al. (2012, in prep.)
for a study of the full (pn+MOS) dataset.

\subsection{Timing}

We performed the analysis of the fast time variability of NGC~5408 X--1 in the 1-8\,keV energy range. The (1-8\,keV) pn count 
rate is $0.2-0.3\,{\rm cts}\,{\rm s}^{-1}$ on average. We excluded from the analysis 
the 8-10\,keV energy band because of the lack of signal and we ignored the 0.3-1\,keV energy range because variability is suppressed in this energy range 
for NGC~5408 X--1 (\cite{middleton11}). We produced light curves with a time bin of 0.5\,s. We used custom software under MATLAB to perform Fast Fourier 
Transforms (FFT) and produce the PDS. A total of 2048 points of the 
light curve were used to perform the FFT. The PDS were normalized according to \cite{leahy83} and converted to 
square fractional root mean square deviation (rms; \cite{belloni90}). PDS fitting was carried out with the standard XSPEC fitting package 
by using a one-to-one energy-frequency conversion and a unit response.

We fitted the PDS with a model constituted by a broad Lorentzian centred at zero frequency plus a constant (expected to be 2
in the Leahy normalization) to account for the Poisson noise. This resulted to 
be a good fit in all the observations, except for the second observation, in which residuals in the form of a QPO are 
visible around $0.01$\,Hz. The integrated fractional rms in the 0.005-1\,Hz band for each observation is reported in Tab.\,\ref{tab1}.

\subsection{Spectroscopy}

We produced background-subtracted energy spectra in the 0.3-10\,keV energy range, with a background
taken from an annulus around the target. Because our goal is to study the intrinsic emission of the ULX, 
in this way we expected to minimize the diffuse emission from the galaxy. Nevertheless, emission lines 
are present in the soft X-rays (0.3-1\,keV) energy range, which may indicate
the presence of residual diffuse emission from the galaxy. We considered the model component {\tt apec} for the description of
this diffuse emission plus emission from the inner disc ({\tt diskpn} model component, 
with corrections for the inner disc emission close to the black hole),
using the {\tt tbabs} absorption model component. To fit the high-energy emission, we considered model components allowing 
for high-energy curvature, i.e. either a powerlaw with a
high-energy cut-off {\tt cutoffpl} or {\tt compTT}, the latter with the temperature for the input soft photons tied to the temperature
of the inner part of the disc. A variable multiplicative 
constant factor was used to account for coordinated variability of the spectra. We fitted the spectra simultaneously, in order to
tie the parameters related to the diffuse emission ({\tt apec} model) and the column density components, expected to be constant. 

In the model with {\tt compTT}, the obtained temperature of the electrons in the corona was ${\rm kT}_{\rm e}=2-3$\,keV and the optical depth ${\tau}=4-5$.
In the phenomenological model with {\tt cutoffpl}, the photon indices and the high-energy cut-off found are in the range $1.7-2.0$ and $3-7$\,keV, respectively. 
The two models provided the same statistical description to the data and none is statistically favoured. The most relevant results of this analysis are 
shown in the following and in Tab.\,\ref{tab1}. 

\section{Results and Implications} 

We found that the X-ray emission of NGC~5408 X--1 is highly variable and that this variability changes during the course of the observations
(i.e. the fractional rms changes from ${\approx}30$\% to ${\approx}50$\%). This level of variability is comparable to BHBs in the low/hard state and to 
some extreme Seyfert galaxies (\cite{vaughan03}). The 
best model applied to the energy-spectra has best-fit parameters that closely resemble those obtained from BHBs during the low/hard state 
(${\Gamma}=1.7-2.0$, ${\rm kT}_{\rm in}{\approx}0.17$\,keV, \cite{reis10}), except for the presence of a low-energy cut-off in the spectra 
(${\rm E}=3-7$\,keV), much lower than those observed in BHBs (${\rm E}=50-100$\,keV). This low value for the cut-off can be physically understood 
as a low value of the temperature of the electrons in the corona and has previously reported in the study of a sample of bright ULXs (\cite{gladstone09}). 

\begin{figure*}[b]
\centering
 \includegraphics[bb=0 0 883 606,width=11.9cm,angle=0,clip]{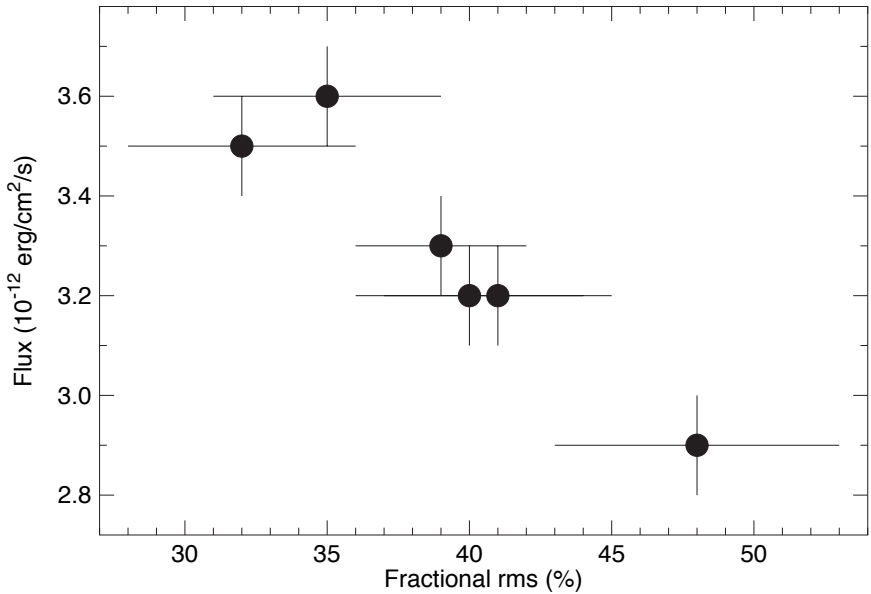}
 \caption{Total unabsorbed flux (in the 0.3-10\,keV energy range) versus the fractional rms of the variability in the 1-8\,keV energy range and 0.005-1\,Hz frequency range.}
\label{fig1}
\end{figure*}

We studied possible correlations of the rms with spectral parameters. We found that there is an anti-correlation of the total
flux with the {\it fractional} rms (see details in the caption of Fig.\,\ref{fig1}), similar to that found in BHBs during the low/hard state 
(\cite{munoz11}). A linear rms-flux correlation in NGC~5408 X-1 has already been reported (\cite{heil10}), but on short time scales
(i.e. within single {\it XMM-Newton} observations; coinciding with our Obs. 1 and 2). Our study shows that there is a
fractional rms-flux anti-correlation in the longer time-scales, in analogy to what observed in the LHS of BHBs (\cite{munoz11}).

\begin{table}
  \begin{center}
  \caption{Results from the timing-spectral study$^1$$^,$$^2$}
\label{tab1}
 {\scriptsize
  \begin{tabular}{cccccc}\hline
{\bf Obs. Number} & {\bf Fractional rms} & {\bf Inner disc temperature} & {\bf Power-law }   & {\bf Energy cut-off }  &  {\bf Unabsorbed flux}   \\
                  &   ($\%$)             &      (keV)                       & {\bf Photon Index} &      (keV)             &     \\ \hline
  1               &  $41{\pm}4$          &   $0.154{\pm}0.003$              & $1.76{\pm}0.11$    &  $3.3{\pm}0.4$         &  $(3.2{\pm}0.1){\times}10^{-12}$   \\ 
  2               &  $48{\pm}5$          &   $0.156{\pm}0.004$              & $1.60{\pm}0.20$    &  $3.1{\pm}0.7$         &  $(2.9{\pm}0.1){\times}10^{-12}$   \\ 
  3               &  $35{\pm}4$          &   $0.155{\pm}0.002$              & $1.87{\pm}0.11$    &  $4.0{\pm}0.6$         &  $(3.6{\pm}0.1){\times}10^{-12}$   \\ 
  4               &  $32{\pm}4$          &   $0.157{\pm}0.002$              & $2.00{\pm}0.10$    &  $5.2{\pm}0.9$         &  $(3.5{\pm}0.1){\times}10^{-12}$   \\ 
  5               &  $39{\pm}3$          &   $0.159{\pm}0.002$              & $1.85{\pm}0.12$    &  $3.9{\pm}0.7$         &  $(3.3{\pm}0.1){\times}10^{-12}$   \\ 
  6               &  $41{\pm}4$          &   $0.162{\pm}0.003$              & $2.10{\pm}0.10$    &  $6.8{\pm}1.9$         &  $(3.2{\pm}0.1){\times}10^{-12}$   \\ 
  \hline
  \end{tabular}
  }
 \end{center}
\vspace{1mm}
 \scriptsize{
 {\it Note:}\\
  $^1$ The results shown are from the model {\tt tbabs(apec+diskpn+cutoffpl)}.  \\
  $^2$ Errors are $1{\sigma}$.
}
\end{table}

\end{document}